\documentclass[manuscript,review=false,anonymous=false]{acmart}
\setcopyright{none}
\makeatletter
\def\@footnotetext#1{}
\def\@copyrightspace{}
\makeatother

\usepackage{tabularx}
\usepackage{booktabs}
\usepackage{array}
\usepackage{enumitem}
\usepackage{placeins}
\usepackage{tcolorbox}
\tcbuselibrary{listings,breakable}
\usepackage{subfig}
\newtcblisting{myverbatim}{
      arc=3mm,
      top=0mm,
      bottom=0mm,
      left=0mm,
      right=0mm,
      boxrule=0.5pt,
      colback=gray!30,
      listing only,
      breakable,
      width=\textwidth, 
        listing options={
    basicstyle=\small\ttfamily,
    breaklines=true,         
    breakatwhitespace=false, 
    columns=fullflexible,    
    keepspaces=true,         
    breakindent=0pt          
  }
}

\begin{document}

\title{To Believe or Not To Believe: Comparing Supporting Information Tools to Aid Human Judgments of AI Veracity}


\author{Jessica Irons}
\email{jessica.irons@csiro.au}
\orcid{0002-0671-5168}
\author{Patrick Cooper}
\email{patrick.cooper@csiro.au}
\author{Necva B{\"o}l{\"u}c{\"u}}
\email{necva.bolucu@csiro.au}
\author{Roelien Timmer}
\email{roelien.timmer@csiro.au}
\author{Huichen Yang}
\email{huichen.yang@csiro.au}
\author{Changhyun Lee}
\email{aiden.lee@csiro.au}
\author{Brian Jin}
\email{brian.jin@csiro.au}
\author{Andreas Duenser}
\email{andreas.duenser@csiro.au}
\author{Stephen Wan}
\email{stephen.wan@csiro.au}
\affiliation{%
  \institution{Commonwealth Scientific \& Industrial Research Organisation}
  \city{Sydney}
  \state{NSW}
  \country{Australia}
}

\renewcommand{\shortauthors}{Irons et al.}

\begin{abstract}
With increasing awareness of the hallucination risks of generative artificial intelligence (AI), we see a growing shift toward providing information tooling to help users determine the veracity of AI-generated answers for themselves. User responsibility for assessing veracity is particularly critical for certain sectors that rely on on-demand, AI-generated data extraction, such as biomedical research and the legal sector. While prior work offers us a variety of ways in which systems can provide such support, there is a lack of empirical evidence on how this information is actually incorporated into the user's decision-making process. Our user study takes a step toward filling this knowledge gap. In the context of a generative AI data extraction tool, we examine the relationship between the type of supporting information (full source text, passage retrieval, and Large Language Model (LLM) explanations) and user behavior in the veracity assessment process, examined through the lens of efficiency, effectiveness, reliance and trust. We find that passage retrieval offers a reasonable compromise between accuracy and speed, with judgments of veracity comparable to using the full source text. LLM explanations, while also enabling rapid assessments, fostered inappropriate reliance and trust on the data extraction AI, such that participants were less likely to detect errors. In additiona, we analyzed the impacts of the complexity of the information need, finding preliminary evidence that inappropriate reliance is worse for complex answers.
We demonstrate how, through rigorous user evaluation, we can better develop systems that allow for effective and responsible human agency in veracity assessment processes.
\end{abstract}





\maketitle

\section{Introduction}

Knowledge management applications increasingly incorporate generative Artificial Intelligence (AI) methods, in which specific details are extracted using from unstructured text.
One particular use of these technologies is to create high-quality, structured datasets compiled from a repository of documents in response to a uniform set of fields or questions. In scientific research, such structured datasets may form the basis of systematic reviews and meta-analyses, and support the development of validated, reusable scientific resources~\cite{wang_evidencebench_2025, schmidt_large_2025, schilling-wilhelmi_text_2025, boorla_catpred_2025, jiang_enzyme_2025, wei_finding_2025}. Off-the-shelf products such as Elicit\footnote{\url{www.elicit.com}} aim to enable researchers to conduct their own dataset extraction by specifying a tabular structure that is filled in using generative AI. 
Industry sectors also rely on such functionality. For example, datasets are created from electronic health records to support evidence-based healthcare~\cite{fornasiere_medical_nodate, shimizu_exploring_2025}. In the legal sector, tools like Nuix\footnote{\url{www.nuix.com}} and Legora\footnote{\url{www.legora.com}} allow users to process very large document datasets and compile tables of extracted details for legal (e-)discovery processes (for overviews of such activities, see~\cite{10.1145/2348283.2348420, 10.1145/3777009}). 

While automated approaches can produce significant time savings~\cite{schmidt_large_2025, pham_design_2025, qian_llm_2025}, this is not without risk.  Data quality issues relating to hallucinations are well-documented, particularly in specialized or niche domains~\cite{wood_role_2010, deng_information_2024, moens_artificial_2025}.  Indeed, there have been well-publicized examples in the legal sector relating to professional misconduct when using AI tools~\citep{artigliereAIhallucinations2026, guardianAIlawyer2025}, prompting the emergence of policies for AI best practice that place the responsibility for the accuracy of generated answers on the AI user.
Similar practices that place the burden of data quality on the user also exist in research and healthcare. 
Ensuring the quality of datasets involves veracity assessment -- ensuring outputs are accurate relative to the source text and making corrections where necessary -- as well as confirming the dataset is complete, internally consistent and appropriately formatted~\cite{wood_role_2010}. This can be a slow and cognitively demanding process, often requiring synthesis across the text or subjective interpretation~\cite{schmidt_large_2025, pham_design_2025, rahman_characterizing_2022, schmidt_data_2025}.

To assist users in this process, commercial systems have adapted to provide supporting information to help users determine the appropriateness of the generated answer.  For example, large language model (LLM) enabled search tools often include links to the original full text (e.g., Perplexity, Google). Elicit provides retrieved passages that align closely with the generated answer (similar to the use of snippets in search engine results).

Current research in this area often focuses on techniques to generate and provide such a facility to users. Comparatively little attention, however, has been paid to how users interpret, trust, and act upon this supporting information. Many supporting information tools aim to reduce users' time and workload by providing short, relevant statements, to spare them searching the full source text. However, efficiency gains must be balanced against the risk of reducing the effectiveness of quality assessment. If key details from the source text are misrepresented or omitted, dataset quality may suffer~\cite{butcher_optimising_2024}. Moreover, as the users' task is to detect AI-generated errors, it is important they maintain a critical stance towards the generated outputs. This can be challenging given the well-documented tendency for over-reliance on automated systems~\cite{parasuraman_humans_1997, simkute2025ironies}, which can lead to inadequate validation. Supporting information, then, should seek to foster appropriate levels of trust and not exacerbate over-reliance.

Consequently, we performed a user study to address this gap. Specifically, in the context of a dataset \textit{veracity assessment} task, users were required to evaluate the correctness of generated details in a dataset relative to the original source.
We assessed two automated forms of supporting information assistance: (1) a BM25-based passage retrieval condition (TopK), where top three retrieved paragraphs were provided; and (2) an LLM-based condition (LLM), where an LLM generated a statement about whether the answer was supported or not by the source material. These were compared against a baseline, no automated support condition (PDF), where participants verified generated answers by referring directly to the source PDF document. We assessed their impact on user task performance metrics (efficiency, accuracy, confidence and acceptance of the AI output), as well as subjective measures of trust, workload and basic usability. 

We evaluated the following research questions: 

\begin{itemize}
    \item \textbf{RQ1.} Do automated supporting information tools (TopK and LLM) reduce the time and perceived workload required for veracity assessment compared with no automated support (PDF)? 
    \item \textbf{RQ2.} Do automated supporting information tools (TopK and LLM) reduce the effectiveness of users' veracity assessments compared with no automated support (PDF)?
    \item \textbf{RQ3.} Does supporting information type impact users' reliance on and trust in AI-generated outputs? 
\end{itemize}

In short, we found that both tooling options---BM25 passage retrieval (TopK) and LLM-generated explanations (LLM)---improved users' efficiency relative to no tooling (PDF), while maintaining high user confidence. However, we found contrasting effects on reliance: LLM explanations led to greater reliance on the generated answers, which impaired users' detection of incorrect generated answers. Overall, TopK passage retrieval was a good compromise for efficiency, accuracy, and usability in simple verification tasks. For more complex verification tasks, directing the participant to the source PDF in its entirety may support higher accuracy.

\section{Background and Approach}
\label{sec:background}

The scenario we consider is one where a human user must verify AI-generated answers. The need for such vetting is underscored by prior work identifying issues that affect automated dataset generation, such as hallucinations (fabricated information), misinterpretation of technical concepts, formatting mismatches and inconsistencies across the data~\cite{ghosh_toward_2024, helms_andersen_using_2025, bianchi_data_2025}.

We note that there is a related area of research: automatic claim verification or fact checking (for example,~\citep{adair_progress_2017, iqbal-etal-2024-openfactcheck, xie-etal-2025-fire, dmonte_claim_2025}).  Tools such as FactScore~\cite{min-etal-2023-factscore}, AlignScore~\cite{zha2023alignscore} and FactCC~\cite{kryscinski2020evaluating} have emerged to \textit{automatically} assess the veracity of the generated text, which can contain multiple claims---albeit to varying levels of success.\footnote{Indeed, automatic metrics for assessing text quality build on earlier NIST evaluation methodology for evaluating information extraction systems, e.g.,~\cite{voorhees_variations_1998, 10.1145/1233912.1233913}.}
In addition to research on developing metrics, others have proposed evaluation frameworks (e.g.,~\cite{iqbal-etal-2024-openfactcheck, es2024ragas}) and datasets~\cite{thorne_fever_2018, wanner2025claimsequalclaimsequal}).

Here, our focus is on the human-centric task of vetting the AI-generated results and how different types of \textit{supporting information} can play a role in decisions on output veracity, particularly for sectors where policy demands human oversight for data quality.  

\begin{figure*}
    \centering
    \includegraphics[width=0.9\textwidth]{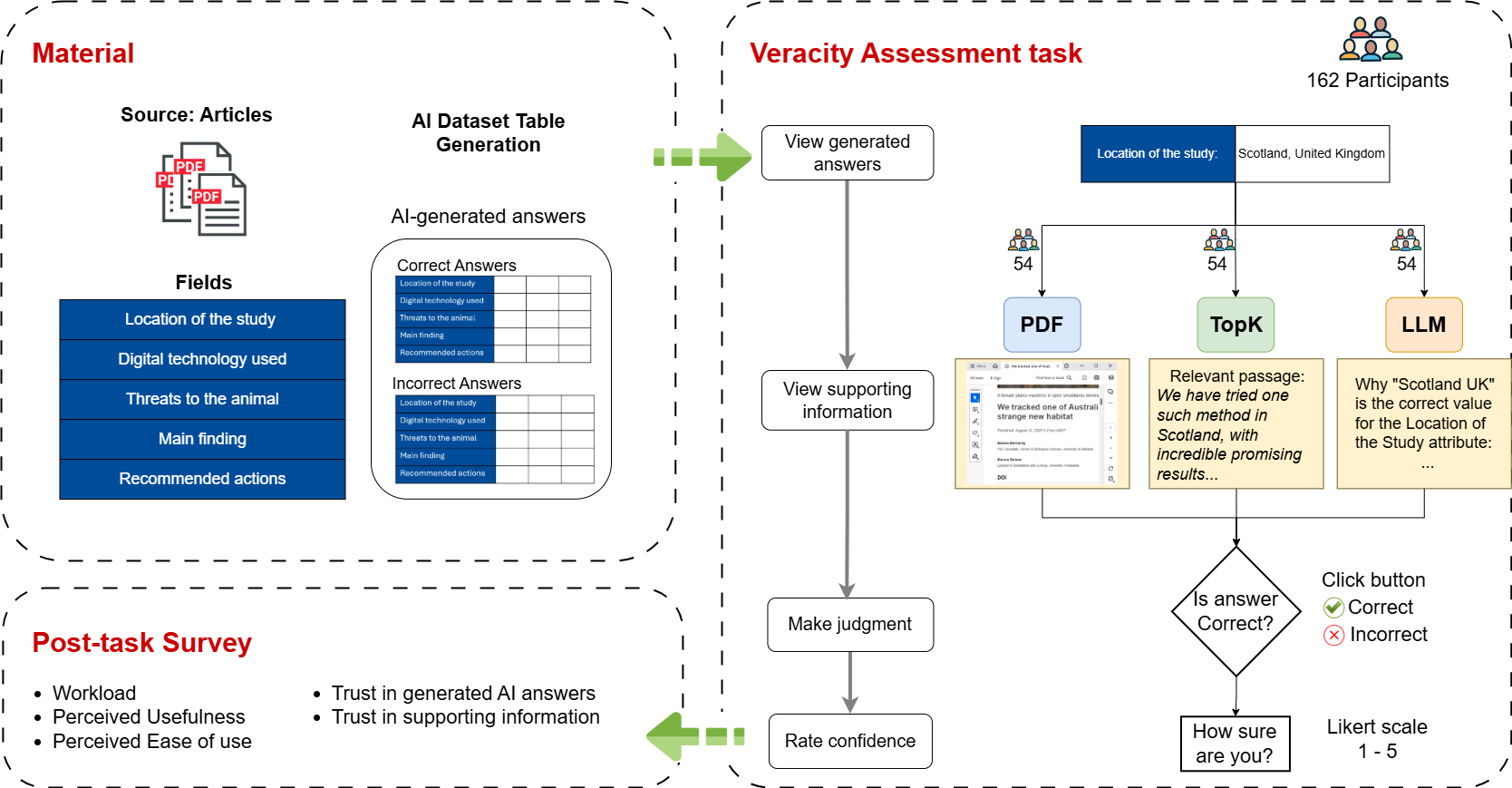}
    \caption{Overview of the user study veracity assessment task} 
    \label{fig:architecture}
\end{figure*}

\subsection{Approaches for presenting supporting information}

In the introduction, we highlighted several examples of how some existing commercial systems use a variety of techniques to help users verify generated answers. As a baseline condition, we use the case where the user is given only the full source text from which the answer was extracted, without further support to aid verification. This is commonly applied by LLM-enhanced search engines that provide links to the full text of a reference document.  

Another approach, used by Elicit and other dataset extraction studies~\cite{butcher_optimising_2024}, is to present a series of extracted texts as supporting evidence.
To retrieve supporting passages, there is a growing body of research focusing on algorithms for the task. For example, in recent years, shared tasks such as Context24~\cite{chan-etal-2024-overview} have helped galvanize efforts to propose techniques for identifying passages of scientific writing, tables and figures to support claims.  

In the Context24 evaluation, variants of standard search engine techniques (e.g., BM25~\cite{10.1561/1500000019}) applied to passage retrieval were shown to represent strong approaches~\cite{bolucu-etal-2024-csiro}, consistent with similar earlier findings on the efficacy of passage retrieval~\citep{fan_generating_2020}. 

The research area of \textit{contextualization} is a closely-related task to activities like Context24.  Here, the aim is to identify additional information from source documentation that is required to properly interpret a claim, which, if taken out of context, could be misunderstood~\cite{10.1145/1555400.1555410, deng-etal-2024-document,10.1145/3077136.3080740}. When this search for extra context relates to evidence, the retrieved information can help assess factual accuracy~\cite{wang_evidencebench_2025, thorne_fever_2018, zeiser_owning_2024}. 
Such work again highlights the potential utility of passage retrieval methods in assessing veracity.

In addition to passage retrieval approaches, LLMs have been used to generate explanations of how the generated answer is (or is not) correct.  The simplest approach is to use prompt-based LLM methods to identify passages of text from the reference that best match the system answer~\cite{wei_finding_2025, zeiser_owning_2024} or explain how the system answer is supported by the reference~\cite{fornasiere_medical_nodate, le_guellec_performance_2024}.
Others have used LLMs to synthesize an explanation given passages retrieved in a prior step (e.g.,~\cite{rahman_characterizing_2022, qian_llm_2025}). 
This LLM information can be produced in an offline process or created on-demand within an ongoing LLM dialogue~\cite{qian_llm_2025} or an interactive search facility~\cite{fan_generating_2020}.

In this user study, we use three conditions broadly representing the spectrum of prior approaches, namely: (1) the full PDF source text (a baseline representing how users would perform the task without additional support), (2) a ``TopK'' passage retrieval approach implemented using BM25 (given prior evidence for its strong performance for this task~\cite{chan-etal-2024-overview}), and (3) an LLM-generated explanation, with supporting evidence, of whether the system answer is supported by the reference~\cite{zeiser_owning_2024}).

\subsection{User Research on Veracity Assessment}
With the data quality issues arising from generative AI, research on how users utilize system outputs is needed. To date, few user studies have compared the impact of different forms of supporting information tooling on how users assess the veracity of AI outputs. There is some evidence that passage retrieval methods improve assessment speed compared with full text search~\citep{butcher_optimising_2024}. However, participants were more prone to missing key information (lower recall), suggesting that relying on an excerpt rather than the full text may compromise completeness (i.e., ensuring all relevant data is captured), a key dimension of dataset quality~\cite{wood_role_2010, koutsiana_knowledge_2024}. LLM-supported explanations have also been shown to enable efficient fact-checking~\cite{si_large_2024}. Qualitative evaluations of dataset assessment show high user satisfaction with LLM-based explanations, particularly when evaluating answers requires more contextual information~\cite{pham_design_2025, qian_llm_2025}. 

Beyond measuring task performance, user research focuses increasingly on trust and reliance, factors that strongly influence how users engage with and adopt AI systems (for related work, see trust in AI machine translation~\cite{martindale-carpuat-2018-fluency, 10.1145/3351095.3372852, 10.1145/3630106.3658941, xiao_toward_2025} ). Trust refers to a belief or attitude toward an AI system, while reliance is a behavioral response reflecting a user's acceptance or rejection of its outputs~\cite{mcgrath_collaborative_2025}. Trust is most effective when appropriately calibrated to the system’s performance and the user’s goals~\cite{lee_trust_2004, mcgrath_collaborative_2025, jacovi_formalizing_2021}. Under-trust results in \textit{disuse}, rejecting system outputs even when they are beneficial, while over-trust leads to \textit{misuse}, accepting the outputs even when they are incorrect~\cite{parasuraman_humans_1997}. 

Appropriate levels of trust and reliance, which are essential for detecting AI errors in veracity assessment, may be influenced by the form of supporting information. Recent work highlights the risks associated with LLM content, which can appear highly persuasive \cite{mcgrath_users_2024}. For example, in a claim verification setting, LLM‑generated explanations were shown to induce over‑reliance, with users more likely to accept false claims when supported by LLM evidence~\citep{si_large_2024}.

\begin{figure*}[t]
\centering
\includegraphics[width=0.9\textwidth]{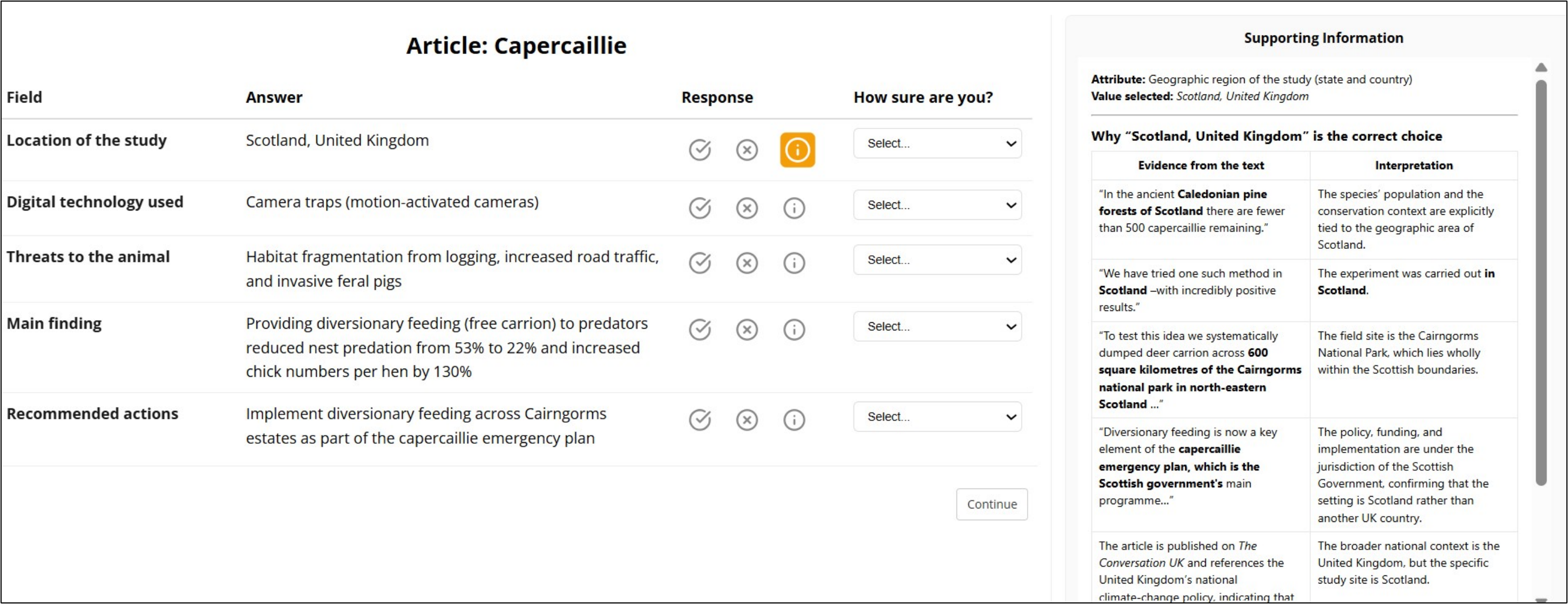}
\caption{Task interface showing one example with LLM supporting information.}
\label{fig:table_review}
\end{figure*}

In the context of dataset veracity assessments, limited empirical evidence exists comparing user performance across these different support tools in a systematic manner. 
Here we take a controlled approach where participants complete a dataset veracity assessment task with one form of supporting information (LLM explanations, passage retrieval or full source text) and incorrect answers are introduced into the dataset in a controlled manner (as in e.g.,~\cite{martindale-carpuat-2018-fluency}). In RQ 1, we evaluated whether supporting information (TopK and LLM) tools reduce the burden on users, by assessing the time taken to perform veracity assessment on AI-generated answers, and users' subjective rating of their workload and usability. RQ 2 compared the impact on the effectiveness of their performance, looking at accuracy (the ability to judge whether AI-generated answers were correct or incorrect) and users' self-reported confidence in their judgment. Finally, RQ 3 evaluated whether supporting information influences trust and reliance on the AI-generated answers. Trust was assessed via a self-report scale, and reliance was measured by the rate at which participants accepted generated answers as correct. 

\section{User Study Method}
\subsection{Task overview and design}
The task required participants to review structured datasets created from science-themed news articles written for a general public readership. Using a generative AI system, we extracted answers for five predefined fields (nominalizations of questions) that were representative of key information from the article (e.g., Location of the study, Main finding).
To evaluate participants' veracity assessment in a controlled way, the correctness of AI answers was systematically manipulated such that, on average, 30\% of answers presented to participants were incorrect. 

The participants’ task was to decide whether each answer was \textit{correct} or \textit{incorrect} using the supplied supporting information. They were randomly assigned to one of three conditions: (\textit{PDF}) original PDF without any additional tooling; (\textit{TopK}; $K=3$) the three most relevant passages retrieved using a BM25-based ranking model; and (\textit{LLM}) an LLM explanation intended to justify the extracted answer. Task performance was assessed using response time to make each veracity decision, accuracy of the veracity decision, self-reported user confidence in their judgment; and percent acceptance, capturing how frequently users accepted or rejected the AI-generated extractions.
Participants also completed a questionnaire on cognitive workload, trust, usefulness, and ease of use of the supporting information provided. The workflow of the task is given in Figure~\ref{fig:architecture}.

\subsection{Materials}

All AI-generated answers and supporting information were generated in advance to ensure consistency across participants. 

\subsubsection{Source texts}
Sources were three scientific articles of reasonable length~\cite{gerhardy_we_2025, lindenmayer_decades_2025, sutherland_surprisingly_2025}, each focusing on conservation efforts for a threatened animal species and published on \textit{The Conversation} website (\url{https://theconversation.com/au}). These articles were selected because they present scientific research in a format accessible to non-expert participants. We intended that the specificity of the details would reduce the familiarity of these topics with our participants, so that they would need to rely on the supporting information for the veracity assessment.

\begin{table*}[th]
\centering
\caption{Examples of correct and incorrect answers used in the experiment.}
\renewcommand{\arraystretch}{1.2}
\begin{tabularx}{\textwidth}{p{3cm} X X}
\toprule
\bf Field & \bf Correct answer & \bf Incorrect answer \\
\midrule
Location of the study &
Scottland, United Kingdom. &
Alberta, Canada. \\
Digital technology used& 
Camera traps (motion-activated cameras). & Unmanned aerial vehicles (UAVs) equipped with thermal-imaging cameras. \\
Threats to the animal &
Predation by pine martens and other carnivores, climate-change induced shifts in spring weather, and resulting loss of eggs and chicks. & Habitat fragmentation from logging, increased road traffic, and invasive feral pigs. \\
Main finding &
Providing diversionary feeding (free carrion) to predators reduced nest predation from 53\% to 22\% and increased chick numbers per hen by 130\%. &
Diversionary feeding led to a 35\% rise in capercaillie breeding success and a 70\% drop in predator sightings. \\
Recommended actions & 
Implement diversionary feeding across Cairngorms estates as part of the capercaillie emergency plan. & Install seasonal feeding stations across Cairngorms estates; restrict hunting permits for predators, and launch a public education campaign on coexistence of capercaillie and pine martens.  \\
\bottomrule
\end{tabularx}
\label{tab:sample_dataset}
\end{table*}

\subsubsection{AI-generated answers}
Our generation pipeline was as follows. Source articles in PDF format were first converted into text using a tool that contains a PDF parser, GROBID~\cite{agosti_grobid_2009} and processed additionally with a table/figure extraction tool. The machine-readable text (from the PDF) and the predefined fields (Location of the study; Digital technology used; Threats to the animal; and Main finding and Recommended actions) were incorporated into a prompt requesting generated answers in a structured format. This prompt was submitted to  GPT-OSS:20B~\cite{openai_gpt-oss-120b_2025}, which generated candidate answers for each field. These fields were chosen to reflect information commonly of interest in scientific review and data curation tasks, while spanning varying levels of semantic complexity from factual to interpretive.  For consistency across study materials, we included only fields for which a valid answer was present in all three articles (as confirmed by one of the authors). 

Evaluation of the generated answers by one author indicated that, in this case, all generated answers were accurate.\footnote{We believe that because the text was written in a style intended for a general public readership and given the relative stylistic similarity to LLM training data from news sources and Wikipedia, LLMs would perform well with such texts.} Minor edits were made to make four of the fifteen answers more concise, without changing their meaning. Therefore, to ensure we had incorrect answers for the veracity assessment task, we artificially created corresponding incorrect answers for each field. We used the same prompt template, adjusting the original prompt by adding ``However, for the purpose of this task, you MUST provide **plausible but incorrect answers**'': 

\begin{myverbatim}
Please, extract {field_1, ..., field_5} from the given article. However, for the purpose of this task, you MUST provide **plausible but incorrect answers**. The output should still be in a list of JSON dictionaries following the given Dictionary Key Mapping, but the answers should NOT be correct according to the article.
[Given Article Start]
{ARTICLE}
[Given Article End]
Remember: the JSON structure must be valid, but the data inside should be wrong in a realistic way.
\end{myverbatim}

All incorrect answers were checked by one author and minor edits were made to ensure that their format and level of specificity closely matched those of correct answers. This was done to prevent participants from distinguishing incorrect answers based on superficial cues alone. An example of correct and incorrect answers used in the experiment is provided in Table~\ref{tab:sample_dataset}.

\subsubsection{Supporting information}
As described in Section~\ref{sec:background}, supporting information was generated using methods representative of those commonly employed in AI information systems. In the PDF condition, supporting information comprised the full article in PDF format, with standard PDF interaction functionality available to participants (e.g., search, scroll). 

In the TopK condition, the supporting information was three paragraphs from the article text with the highest lexical relevance to the generated answer and its corresponding field. Paragraph relevance was computed using the BM25 retrieval model, where the query was constructed from the generated answer and its corresponding field, and the full article text served as the retrieval corpus. We selected three paragraphs as a heuristic decision that allows for limited over-retrieval (i.e., more than a single paragraph) without avoiding excessive cognitive load. Separate sets of supporting paragraphs were created for each correct and incorrect answer, and for each of the 15 answers (3 articles X 5 fields). 

In the LLM condition, we used GPT-OSS:20b (thinking=low) to generate a natural-language explanation for each AI-generated answer and used its output as the supporting information. The model was prompted as follows:

\begin{myverbatim}
You are provided with the following context from a document. Based on this information, explain your reasoning for the attribute '{field label}' with answer '{answer}'. Be clear and justify why this answer is appropriate.

[Context]
{ARTICLE}
\end{myverbatim}

In this way, the LLM was explicitly instructed to take a stance on the veracity of the AI-generated answer with respect to the provided content (article text). Although the prompt asked the LLM to justify the answer, for incorrect responses, the output sometimes produced an explanation for why the answer was not correct. These outputs were retained as-is and presented to participants.

\subsubsection{Survey measures}\label{sssec:subjectivemeasures}
A post-task survey (see Table~\ref{tab:survey}) was constructed to assess the following subjective measures: \textit{Cognitive Workload, Trust in the AI Answers, Trust in the Supporting Information, Usefulness of the Supporting Information and Ease of Use of the Supporting Information}. 

Trust in both the AI Answers and in the supporting information was assessed using the 3-item Short Trust in Automation Scale (S-TIAS)~\cite{mcgrath_measuring_2025} and rated on a 7-point Likert scale. Cognitive workload was assessed using the NASA Task Load Index (TLX)~\cite{hart_development_1988}. For brevity, we included four items (Mental demand, Performance, Effort, and Frustration) and excluded two items not relevant to the task (Physical demand, temporal demand). Ratings were given on a 21-point sliding scale from very low (0) to very high workload (20). Usefulness and Ease of Use were assessed using adapted scales from~\citet{davis_perceived_1989}. As not all items were relevant to the task at hand (e.g., related to the participants’ everyday work), we selected 4 items from Usefulness and 3 from Ease of Use. Participants rated the items on a scale from 1 (``almost never'') to 5 (``almost always'').

\begin{table*}[ht]
\centering
\caption{Subjective measure survey items}
\renewcommand{\arraystretch}{1.2}
\begin{tabularx}{\textwidth}{p{5cm} X}
\toprule
\bf Subjective measure (\textit{Scale}) & \bf Survey items \\
\midrule

\parbox[t]{5cm}{Workload\\ \textit{NASA-TLX}~\cite{hart_development_1988}} &
\begin{itemize}
    \item How mentally demanding was the task? 
    \item How successful were you in accomplishing what you were asked to do? 
    \item How hard did you have to work to accomplish your level of performance? 
    \item How insecure, discouraged, irritated, stressed, and annoyed were you? 
\end{itemize}
\\

\parbox[t]{5cm}{Trust in AI answers \\ (\textit{S-TIAS})~\citep{mcgrath_measuring_2025}} & 
\begin{itemize}
\item I could trust the AI tool 
\item I was confident in the AI tool 
\item The AI tool was reliable 
\end{itemize}
\\

\parbox[t]{5cm}{Trust in supporting information \\ (\textit{S-TIAS})~\citep{mcgrath_measuring_2025}}  &  
\begin{itemize}
\item I could trust the supporting information 
\item I was confident in the supporting information 
\item The supporting information was reliable 

\end{itemize}
\\

\parbox[t]{5cm}{Usefulness of the supporting information  \\ (\textit{Perceived Usefulness})~\citep{davis_perceived_1989}} & 
\begin{itemize}
\item Using the supporting information improved my performance 
\item Using the supporting information increased my productivity 
\item Using the supporting information helped me accomplish the task more quickly 
\item I found the supporting information useful 
\end{itemize}
\\
\parbox[t]{5cm}{Ease of use of the supporting information \\ (\textit{Perceived Ease of Use})~\citep{davis_perceived_1989}} & 
\begin{itemize}
\item Learning to use the supporting information was easy 
\item It was easy for me to become skillful at using the supporting information 
\item The supporting information was easy to navigate 

\end{itemize}
\\
\bottomrule
\end{tabularx}
\label{tab:survey}
\vspace{2mm}
\footnotesize
\textit{Note.} NASA-TLX = NASA Task Load Index. S-TIAS = Short trust in automation scale.
\end{table*}

\subsection{Task interface and procedure}
The veracity assessment interface was implemented using jsPsych v7.3 and delivered via an in-house server running JATOS. The interface displayed dataset tables containing AI-generated answers, alongside interactive controls for validation and access to supporting information. 

At the beginning of the task, participants were informed that the dataset had been generated by an AI-system, and their task was to judge the correctness of the generated answers. For each AI-generated answer, participants could view supporting information in a side panel (see Figure~\ref{fig:table_review}) by clicking an information button next to the answer. 

In the PDF group, participants were shown the entire article in PDF format and informed that they may use zoom and search functionality is desired. In the TopK and LLM groups, supporting information was presented in a text box; the full article was not available. In the instructions, TopK supporting information was described as ``You will see a few relevant paragraphs from the article'', while the LLM explanation was described as ``A different AI will check the answer and explain why it thinks the answer is correct or not''. 

On a given article, participants could evaluate the five generated answers in any order. After responding to each answer (selecting the tick or cross button), they were prompted to rate their confidence in the accuracy of their judgment (``How sure are you?'') on a Likert scale from 1 (not at all) to 5 (very sure). On completion of all five answers, participants clicked \textit{Continue} to move to the next article. At the end of the task, participants completed the post-task survey assessing subjective measures (see Section~\ref{sssec:subjectivemeasures}).

\subsubsection{Subjective measures}\label{sssec:subjectivemeasures}
The survey (see Table~\ref{tab:survey}) was conducted to assess the following subjective factors: \textit{Cognitive Workload, Trust in the AI Answers, Trust in the Supporting Information, Usefulness of the Supporting Information and Ease of Use of the Supporting Information}. 

Trust in both the AI Answers and in the supporting information was assessed using the 3-item Short Trust in Automation Scale (S-TIAS)~\cite{mcgrath_measuring_2025} and rated on a 7-point Likert scale. Cognitive workload was assessed using the NASA Task Load Index (TLX)~\cite{hart_development_1988}. For brevity, we included four items (Mental demand, Performance, Effort, and Frustration) and excluded two items not relevant to the task (Physical demand, temporal demand). Ratings were given on a 21-point sliding scale from very low (0) to very high workload (20). Usefulness and Ease of Use were assessed using adapted scales from~\citet{davis_perceived_1989}. As not all items were relevant to the task at hand (e.g., related to the participants’ everyday work), we selected 4 items from Usefulness and 3 from Ease of Use. Participants rated the items on a scale from 1 (``almost never'') to 5 (``almost always'').

\subsection{Participants}
Participants were recruited through the crowdsourcing platform Prolific~\cite{palan_prolificacsubject_2018}\footnote{\url{https://www.prolific.com/}}. Based on a power analysis (medium effect size; power = 80\%), we recruited 54 per group (162 total; 66 female, 1 non-binary, and 95 male; mean age = 40.57; all US-based. Participants gave informed consent prior to the task and were compensated at Prolific's recommended rate (\$12USD/hour). The experiment took on average 17 mins. All methods were approved by the 
CSIRO Social and Interdisciplinary Science Human Research Ethics Committee.

\begin{table}[th]
\centering
\caption{Mean scores on task performance and subjective measures.}
\label{tab:scores}
\begin{tabular}{l c c c l}
\toprule
 & \multicolumn{3}{c}{Supporting Info.} & \\
\cmidrule(lr){2-4}
Measure & PDF & TopK & LLM & ANOVA Results \\
\midrule
\multicolumn{5}{l}{\textit{Task performance}} \\
\textbf{RT (sec)} & \textbf{45.8} & \textbf{33.0} & \textbf{25.4} & \footnotesize \textit{F}(2,159)=15.52, \textit{p}<.001, $\eta_p^2$=.16 \\
Accuracy (\%) & 85.7 & 83.1 & 80.8 & \footnotesize \textit{F}(2,159)=2.52, \textit{p}=.08 \\
Confidence & 4.1 & 4.2 & 4.2 & \footnotesize \textit{F}(2,159)=0.51, \textit{p}=.60 \\
\textbf{Acceptance (\%)} & \textbf{71.5} & \textbf{64.7} & \textbf{86.5} & \footnotesize \textit{F}(2,159)=45.48, \textit{p}<.001, $\eta_p^2$=.36 \\
\midrule
\multicolumn{5}{l}{\textit{Subjective measures}} \\
\textbf{Trust AI Ans.} & \textbf{4.1} & \textbf{4.3} & \textbf{5.4} & \footnotesize \textit{F} (2,159)=14.27, \textit{p}<.001, $\eta_p^2$=.15 \\
Trust Sup. Inf. & 5.7 & 5.5 & 5.7 & \footnotesize \textit{F}(2,159)=0.69, \textit{p}=.51 \\
\textbf{Workload} & \textbf{11.6} & \textbf{9.7} & \textbf{9.5} & \footnotesize \textit{F}(2,159)=8.38, \textit{p}<.001, $\eta_p^2$=.10 \\
\textbf{Usefulness} & \textbf{4.1} & \textbf{4.3} & \textbf{4.5} & \footnotesize \textit{F}(2,159)=5.23, \textit{p}=.006, $\eta_p^2$=.06 \\
\textbf{Ease of use} & \textbf{3.6} & \textbf{4.2} & \textbf{4.2} & \footnotesize \textit{F}(2,159)=7.86, \textit{p}<.001, $\eta_p^2$=.09 \\
\bottomrule
\end{tabular}

\vspace{2mm}
\footnotesize
\textit{Note.} ANOVA results report main effects across supporting information groups.
Bold indicates a significant effect. Acceptance = percent acceptance of the AI-generated answers.
Trust AI Ans. = Trust in the AI generated answers. Trust Sup. Inf = Trust in the supporting information.
\end{table}

\subsection{Data Analysis}\label{ssec:data_analysis}
Data collected on the task were screened to exclude responses on articles where participants appeared not to use the supporting information. In the PDF condition, an article was excluded if the participant did not open the PDF supporting information at least once (since the same information applied to all five answers, opening it once per article was sufficient). In the TopK and LLM conditions, articles were excluded if participants accessed supporting information for fewer than four answers. This threshold (4 instead of 5) accounts for cases where multiple answers could be found in one piece of supporting information. Additionally, any articles completed in under 30 seconds were removed. Participants with two or more articles excluded were dropped from the analysis, and additional participants were recruited to maintain a sample size of 54 per group.

Participants' veracity assessment performance was calculated for each of the 15 responses (5 answers x 3 articles). Accuracy in veracity assessment was measured by directly comparing the participant’s judgment (correct/incorrect) to the ground truth. Judgment response time (RT) reflected the time elapsed between events (i.e., button presses or page loads) and was analyzed for correct responses only. AI acceptance was the percentage of responses in which participants accepted the generated answers to be correct. Mean scores were also calculated for the five subjective measure sub-scales. 
Data were analyzed using between-subjects and mixed ANOVAs. Follow-up simple effects tests were conducted using the Holm-Bonferroni correction for family-wise error rate~\cite{holm_simple_1979}; reported p-values have been adjusted accordingly.

\section{User Study Results}

\paragraph{RQ1: Do automated supporting information tools (TopK and LLM) reduce the time and perceived workload required for veracity assessment compared with no automated support (PDF)?}
As shown in Table~\ref{tab:scores}, RTs differed across the three groups. RTs were slower for PDF compared with TopK (\textit{t}(106) = 3.24, \textit{p} = .003, \textit{d} = .62) and LLM (\textit{t}(106) = 5.11, \textit{p} < .001, \textit{d} = .98), consistent with the prediction that supporting information tools reduce the time required for veracity assessments. Comparing the two tooling options, we found that judgments faster in the LLM than TopK condition (\textit{t}(106) = 2.45, \textit{p} = .015, \textit{d} = .47).

Subjective measures of both workload and usability also revealed poorer outcomes overall in the PDF condition (Table~\ref{tab:scores}). Participants in this group reported higher workload than those in TopK (\textit{t}(106) = 3.43, \textit{p} = .002, \textit{d} = .66) and LLM (\textit{t}(106) = 3.65, \textit{p} = .001, \textit{d} = .70; no significant difference between TopK and LLM (\textit{t}(106) = 0.44, \textit{p} = .66). Similarly, Ease of Use of the supporting information was lower for PDF than TopK (\textit{t}(106) = 3.41, \textit{p} = .003, \textit{d} = .66) and LLM (\textit{t}(106) = 3.19, \textit{p} = .004, \textit{d} = .61; TopK vs LLM not significant, \textit{t}(106) = .04, \textit{p} = .97). Usefulness of the supporting information was lower in PDF than LLM (\textit{t}(106) = 3.30, \textit{p} = .004, \textit{d} = .63), while TopK did not differ significantly from either group (vs. PDF, \textit{t}(106) = 1.52, \textit{p} = .18; vs. LLM, \textit{t}(106) = 1.70, \textit{p} = .18).

\begin{figure*}[ht]
\centering
\includegraphics[width=1\textwidth]{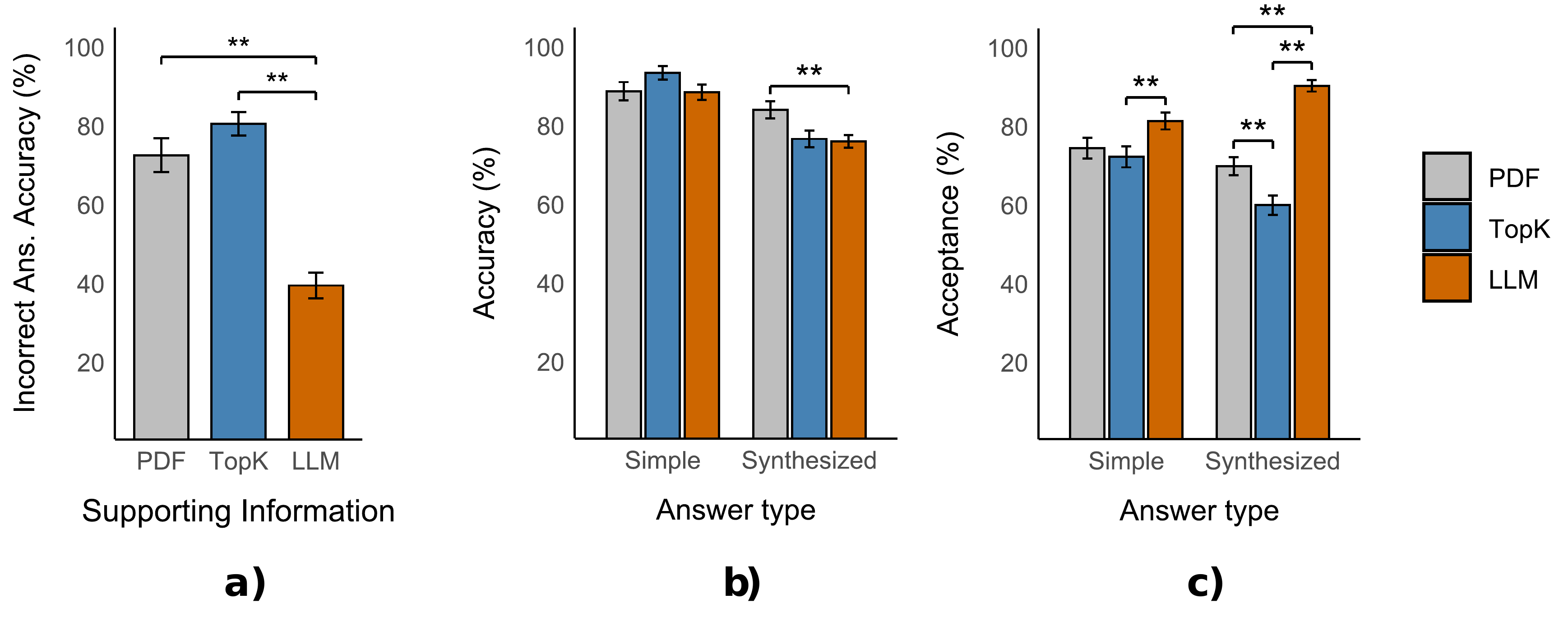}
\caption{ a) Mean accuracy on incorrect answers across the three supporting information groups (PDF, TopK, LLM). b) Mean accuracy and c) mean acceptance of the AI answer across the three supporting information groups (PDF, TopK, LLM) as a function of answer type (simple vs synthesized). ** indicates significant difference between groups. Error bars show the standard error of the mean.}
\label{fig:figAcc}
\end{figure*}

\paragraph{RQ2. Do automated supporting information tools (TopK and LLM) reduce the effectiveness of users' veracity assessments compared with no automated support (PDF)?}
Analysis of overall scores did not support the view that TopK and LLM resulted in lower accuracy or confidence. Mean accuracy did not vary significantly across the groups, nor did participants’ confidence in their judgments (see Table~\ref{tab:scores}). 

\paragraph{RQ3. Does supporting information type impact users' reliance on and trust in AI-generated outputs?}
To assess participants' reliance on the AI-generated answers, we evaluated how frequently they accepted the answer. We found that acceptance differed significantly across groups (Table~\ref{tab:scores}) and was highest in the LLM condition compared with both PDF (\textit{t}(106) = 6.95, \textit{p} = < .001, \textit{d} = 1.34) and TopK (\textit{t}(106) = 9.87, \textit{p} < .001, \textit{d} = 1.90). Acceptance was lowest in TopK (vs. PDF, \textit{t}(106) = 2.59, \textit{p} = .011, \textit{d} = .50). 

Given that we fixed the correctness of the answers at 70\%, an appropriate level of reliance would mean participants should accept the AI approximately 70\% of the time. Acceptance rates were significantly higher than this benchmark in the LLM condition \textit{t}(53) = 14.11, \textit{p} < .001, \textit{d} = 1.92), and significantly lower in TopK (\textit{t}(53) = 2.83, \textit{p} = .01, \textit{d} = 0.38).

Inappropriately high reliance can result in a failure to detect AI errors. Therefore, we compared participants' accuracy on \textit{incorrect} AI-generated answers only. Consistent with this hypothesis, accuracy on incorrect answers differed across groups (\textit{F}(2, 159) = 37.37, \textit{p} < .001, $\eta_p^2$ = .32), with mean scores significantly poorer in the LLM condition compared with both the PDF (\textit{t}(106) = 6.13, \textit{p} < .001, \textit{d} = 1.18) and LLM (\textit{t}(106) = 9.27, \textit{p} < .001, \textit{d} = 1.78); see Figure ~\ref{fig:figAcc}a).

On participants' self-reported trust, we found no differences in how much participants trusted the supporting information itself; however, we did observe differences in how much trust they placed in the \textit{AI generating the dataset answers}. Trust in the AI was significantly higher in the LLM condition over both PDF (\textit{t}(106) = 5.03, \textit{p} < .001, \textit{d} = .97) and TopK (\textit{t}(106) = 4.27, \textit{p} < .001, \textit{d} = .82), with no difference between the latter two (\textit{t}(106) = .74, \textit{p} = . 46). 

A clearer picture of behavior emerges when looking at the relationship between trust and reliance. Across all participants, those who report higher trust in the AI also had higher acceptance of the generated answers (\textit{r} = .57, \textit{t}(106) = 8.78, \textit{p} < .001). Together with the LLM group’s higher scores on both measures, these findings suggest a shared mechanism by which supporting information influenced reliance and trust in the AI together. Trust in the supporting information did not correlate with reliance, \textit{r} = .06, \textit{t} = .78 \textit{p} = .25.

\subsection{Exploratory Analyses: Answer type}

To further explore our findings, we examined whether the effects of supporting information depended on the type of the answer being evaluated. This was motivated by prior qualitative user studies showing that users find validating complex, integrative answers especially challenging~\cite{pham_design_2025}. Short forms of supporting information (TopK and LLM) may lack all necessary information to accurately evaluate answers that require synthesis across greater portions of the text ~\cite{butcher_optimising_2024}. To explore this possibility, we split the data into responses to simple fields (i.e., short, singular answers, specifically Location of the study and Digital technology used) and synthesized fields (i.e., answers comprised either a list or a summary, specifically Threats to the animal, Main finding and Recommended actions).\footnote{We confirmed the number of incorrect answers shown to participants were the same across the two answer types (\textit{p} = .62).} 

Evaluating synthesized answers was indeed more difficult overall: both accuracy (\textit{F}(2, 159) = 59.42, \textit{p} < .001, $\eta_p^2$ = .27) and confidence ratings (\textit{F}(2, 159) = 92.64, \textit{p} < .001, $\eta_p^2$ = .37) were lower. For accuracy, we observed an interaction between answer type and group (see Figure~\ref{fig:figAcc}b; \textit{F}(2, 159) = 5.75, \textit{p} = .004, $\eta_p^2$ = .07). For simple answers, supporting information did not affect accuracy (all comparisons between groups \textit{p}s > .24). For synthesized answers, difference between groups were more apparent. PDF led to significantly higher accuracy than LLM (\textit{t}(106) = 2.97, \textit{p} = .02, \textit{d} = .57). Accuracy was also numerically higher in PDF than TopK, although this difference was not significant after controlling for multiple comparison (\textit{t}(106) = 2.43, \textit{p} = .08). No significant interaction emerged for confidence ratings (\textit{F}(2, 159) = 1.50, \textit{p} = .23).

We also explored whether reliance, as measured by acceptance rates, differed across answer types. Again we observed a significant interaction with supporting information (\textit{F}(2, 159) = 11.55, \textit{p} < .001, $\eta_p^2$ = .13). As shown in Figure ~\ref{fig:figAcc}c, for simple answers there was relatively little variation across groups, with the only significant difference between the LLM and TopK conditions (\textit{t}(106) = 2.66, \textit{p} = .03, \textit{d} = .51; all other \textit{p}s > .08). In contrast, differences between groups were more pronounced for synthesized answers. Acceptance rates were significantly higher for LLM than both PDF (\textit{t}(106) = 7.57, \textit{p} < .001, \textit{d} = 1.46) and TopK (\textit{t}(106) = 10.66, \textit{p} < .001, \textit{d} = 2.05). Additionally, acceptance in the TopK condition was lower than in PDF \textit{t}(106) = 2.96, \textit{p} = .02, \textit{d} = .57.  

Overall, the exploratory analyses indicate that the impact of supporting information is more pronounced for synthesized answers, with LLM-based support increasing acceptance and reducing accuracy on these more complex judgments.

\section{Discussion}

\subsection{Key Findings}

Our findings show that there are effective ways to help users verify AI-generated answers. Presenting either retrieved passages or LLM explanations increased the efficiency of users' veracity assessment without negatively affecting their accuracy or  confidence. Furthermore, both fared favorably regarding perceptions of workload, usefulness, and ease of use. 
Where the two approaches diverged was in their effects on reliance and trust. LLM explanation led to higher levels of reliance and trust in the AI, and TopK to lower levels of reliance, compared with the PDF-only condition. For the LLM condition, this was associated with poorer detection of AI-generated errors. This complements related comparisons of LLMs (using the knowledge represented within the LLM alone) and web search engines for fact-checking claims~\cite{si_large_2024}, which found a similar case of inappropriate reliance on the LLM within that scenario. 

To understand why TopK and LLM produced these different outcomes, we drew on human-automation teaming research evaluating how automating different parts of the user's task impact overall task performance~\cite{endsley_out---loop_1995, parasuraman_model_2000, irons_towards_2025}. Automating \textit{information acquisition} often produces benefits to task performance, whereas automation in \textit{decision‑making} increases new risks, including users becoming less able to detect or recover from errors. This is attributed to increased complacency and a reduction in contextual knowledge, as users become cognitively disengaged from the task and rely more heavily on AI. Our results extend this pattern to veracity assessment in the IR context: supporting information tooling that was limited to acquiring relevant content---TopK---produced generally positive outcomes. When the supporting information also recommended a decision---LLM---inappropriate reliance led to missed errors.

The current findings also highlight that trust in AI can manifest in nuanced and indirect ways, not only in how explanations are perceived but in how users respond to other parts of the system. While participants did not trust the LLM explanation any more than the PDF or TopK (no difference in Trust in the Supporting Information), its presence appeared to confer greater trust in the AI that generated the answers. We also observed that reliance (acceptance) was related to trust in the AI, suggesting a shared mechanism by which supporting information influenced reliance and trust in the AI together. One interpretation is that participants accepted the generated answers more frequently in the LLM condition and therefore inferred greater accuracy of the generative AI system, which in turn led to higher trust. In practice, the effect may be cyclical, with higher trust leading to less critical evaluation, ultimately undermining the effectiveness of veracity assessment.

A potential qualification on the use of supporting‑information tools relates to complexity of the answers. For more complex, synthesized answers (lists or summaries), both TopK and LLM produced numerically lower accuracy than the PDF condition. This finding is intuitive as important information can be omitted from concise forms of supporting information~\cite{butcher_optimising_2024}. Reliance on the AI answer also increased in the LLM condition for synthesized answers. This may imply effort avoidance, with participants preferring to accept the easy AI answer when validating is challenging. In contrast, TopK showed a tendency towards under-reliance. While we can only speculate, it is possible that when the supporting information was insufficient to allow full validation, participants had a preference towards rejecting the AI answer. While work is required to validate these exploratory findings and unpack the underlying mechanisms, they highlight the importance of considering the nature of the extracted information when evaluating supporting information tools. 

\subsection{System Design Implications}

Studies such as ours can help validate the benefits and risks of employing particular evidence-supported tooling in AI applications. Understanding the risks of generative AI (e.g. hallucination) is particularly important given the obligation for researchers and system designers to follow Responsible AI and FAIR principles.  
Developing systems to place the onus on the user to take responsibility for outcomes is an active field of research.  For example, simply indicating the fallibility of AI systems (e.g., by using first-person hedging phrases) has been shown to decrease AI trust and increase user accuracy~\cite{kim_im_2024}.  Our findings here suggest that designing the system to present a composite set of supporting information may be a strategy to mitigate potentially inappropriate reliance on AI results.  Allowing the user flexibility to choose the kind of supporting information mirrors related findings in other language technology settings.  For example,~\citet{xiao_toward_2025} also found an over-reliance on system output in the context of machine translation, even when such translations were not perfect.  The authors note that users have no other choice, suggesting the importance of providing a range of alternatives for users to assess AI output quality.

\subsection{Limitations and Future Work}
This study has several limitations. 
First, for ease of participant recruitment, we relied on a sample from the general public.  This pragmatic step influenced our choice of materials (science communication articles).  As such, participants were not deeply invested in dataset quality (beyond task participation). 
Professional data curators who are highly motivated to produce good-quality datasets and are familiar with the structure of source text may have different preferences for supporting information and be less inclined to accept LLM-generated explanations. While novice curation is not unrealistic (e.g., genomics curation often involves student or citizen science communities~\cite{ejigu_review_2020}), future work should validate these findings with domain experts in real-world settings and text types.

Second, the choice to limit participants to only one form at a time enabled us to investigate the supporting information types in isolation. Depending on the system, however, users may not be restricted to a single information source and may combine multiple sources or switch between them depending on task complexity and personal preferences. Investigating hybrid or adaptive support strategies remains an important direction for future work.

Third, the LLM condition used an explanation style that explicitly recommended a decision. Alternative LLM-based approaches should be explored in future work. Explanations that seek to provide evidence without prescribing a decision may support validation while reducing the risk of over-reliance. For example, contrastive explanations---prompting the LLM to compare two outcomes or provide a case for and against the claim---have shown promise in claim verification tasks and may reduce (though not eliminate) over-reliance~\cite{si_large_2024, bucinca_contrastive_2025}. 

\section{Conclusion}
We describe findings from a user study designed to evaluate the utility of LLM and passage retrieval tools to help users assess the accuracy of AI-generated answers. In particular, we examined performance along dimensions of efficiency, effectiveness, trust in and reliance on AI.  Using data with known errors, the task was to vet generated answers for pre-determined fields (verbalized in as nominalizations) using three kinds of supporting information: the full reference text (PDF), a BM25 passage retrieval tool (TopK), and LLM explanation designed to indicate if the claim was supported or not (LLM). We found that compared with the full reference, the TopK supporting information allowed participants to reach comparable quality levels at a greater speed. Furthermore, while the LLM information also led to quick decision-making, this was accompanied by inappropriately high levels of reliance on the AI capability --- to the detriment of accuracy. We found evidence that complex information needs (i.e., long answers) can exacerbate the inappropriate reliance on LLMs explanations. 
Such findings demonstrate how rigorous user experimentation can shed new light on and help bridge knowledge gaps towards designing AI systems that are based on effective human-AI teaming and support appropriately calibrated human trust and reliance.

\bibliographystyle{ACM-Reference-Format}
\bibliography{SIGIR-user}

\end{document}